\documentclass{iau}
\usepackage{graphicx}
\usepackage{natbib}
\bibpunct{(}{)}{;}{a}{}{,} % to follow the A&A style

\title%[Short title] %% give here short title %%
{Collisions of planetesimals and formation of planets}

\author[R.\ Dvorak et al.]%\& B. Author]   %% give here short author list %%
{Rudolf Dvorak$^1$, %\thanks{email: {\tt rudolf.dvorak@univie.ac.at}},
Thomas I.\ Maindl$^1$, \'Aron S\"uli$^2$, Christoph M.\ Sch\"afer$^3$, Roland Speith$^4$, 
 \and Christoph Burger$^1$}

\affiliation{$^1$Department of Astrophysics, University of Vienna, 
Austria\\email: {\tt rudolf.dvorak@univie.ac.at}\\

$^2$E\"otv\"os University, Department of Astronomy, %H-1518 
Budapest, %Pf.\ 32, 
Hungary\\

$^3$Institut f\"ur Astronomie und Astrophysik, Eberhard Karls Universit\"at T\"ubingen, Germany\\%,
$^4$Physikalisches Institut, Eberhard Karls Universit\"at T\"ubingen, 
Germany}

\pubyear{2015}
\volume{314}  %% insert here IAU Symposium No.
\pagerange{119--126}
\jname{Young Stars \& Planets Near the Sun}
\editors{J. H. Kastner, B. Stelzer, \& S. A. Metchev, eds.}
\begin{document}

\maketitle

\begin{abstract}
%The outcome of numerical simulations of the formation of planets depends a lot on the choice of the initial distribution of planetesimals and planetary embryos after the disappearance of gas in the protoplanetary disk. We take into account that some of these planetesimals of sizes in the order of the mass of the Moon already contain water; the quantity depends on the distance from the Sun --- too close and the bodies are dry, but starting from a distance of about 2\,AU they can contain substantial amounts of water. We show preliminary results of terrestrial planet formation using on one side classical numerical integrations of hundreds of small bodies on CPUs and on the other side --- for comparison reasons --- the results of our GPU code with thousands of small bodies which then merge to larger ones. To be able to determine the outcome of collisions we use our SPH code which shows how water is lost during such events.
We present preliminary results of terrestrial planet formation using on the one hand classical numerical integration of hundreds of small bodies on CPUs and on the other hand---for comparison reasons---the results of our GPU code with thousands of small bodies which then merge to larger ones. To be able to determine the outcome of collision events we use our smooth particle hydrodynamics (SPH) code which tracks how water is lost during such events.
\end{abstract}

\firstsection % if your document starts with a section,

\section{Introduction and scenarios}

The outcome of numerical simulations of planet formation is extremely sensitive to initial conditions like disk surface density, total mass, the initial distribution of planetesimals and planetary embryos after the gas phase in the protoplanetary disk, and the dynamical model. Hence, the resulting planetary systems and even statistical statements are highly biased by the choice of these parameters. While many results on formation of planets and their water content have been achieved, several open questions remain \citep[e.g., ][]{raykok14}.

%We take into account that some of these planetesimals of sizes in the order of the mass of the Moon already contain water; the quantity depends on the distance from the Sun: too close and the bodies are dry, but starting from a distance of about 2\,AU they can contain substantial amounts of water.

%We consider three models: existing gas giants Jupiter and Saturn (model {A}), only Jupiter (model {B}), and without any giant planet (model {C}). While all three models are important for the understanding of extrasolar planetary systems, we present preliminary results for model {A}.

Here, we present preliminary dynamical results for a model including two gas giants, Jupiter and Saturn. By SPH simulation of collision events we can deduce the outcome of collisions in terms of merging/fragmentation and water loss depending on mass and size of the bodies, collision velocity, and impact angle \citep[Fig.~\ref{fig:gpu}b and][]{maisch13,maidvo14b}, which addresses the yet unclear contribution of including water transport probabilities in planet formation simulations \citep[cf.][]{izihag14}.

We take the outcome of the last phases of the Grand Tack scenario \citep{walmor11} as initial conditions for our simulations. Jupiter and Saturn are in their actual positions in the Solar System and we distribute small bodies in the mass range $3\times 10^{-9}\,M_\odot < m < 3\times 10^{-7}\;M_\odot\/$ and semi-major axes $0.4\,\mbox{AU} < a < 2.7\,\mbox{AU}\/$ with small inclinations and small eccentricities. Bodies initially outside 2\,AU are assumed to have a water content of 10 percent.

\section{Preliminary simulation results}

\begin{figure}
        \begin{center}
          {\includegraphics[clip,trim=0 2mm 0 7mm,width=0.45\linewidth]{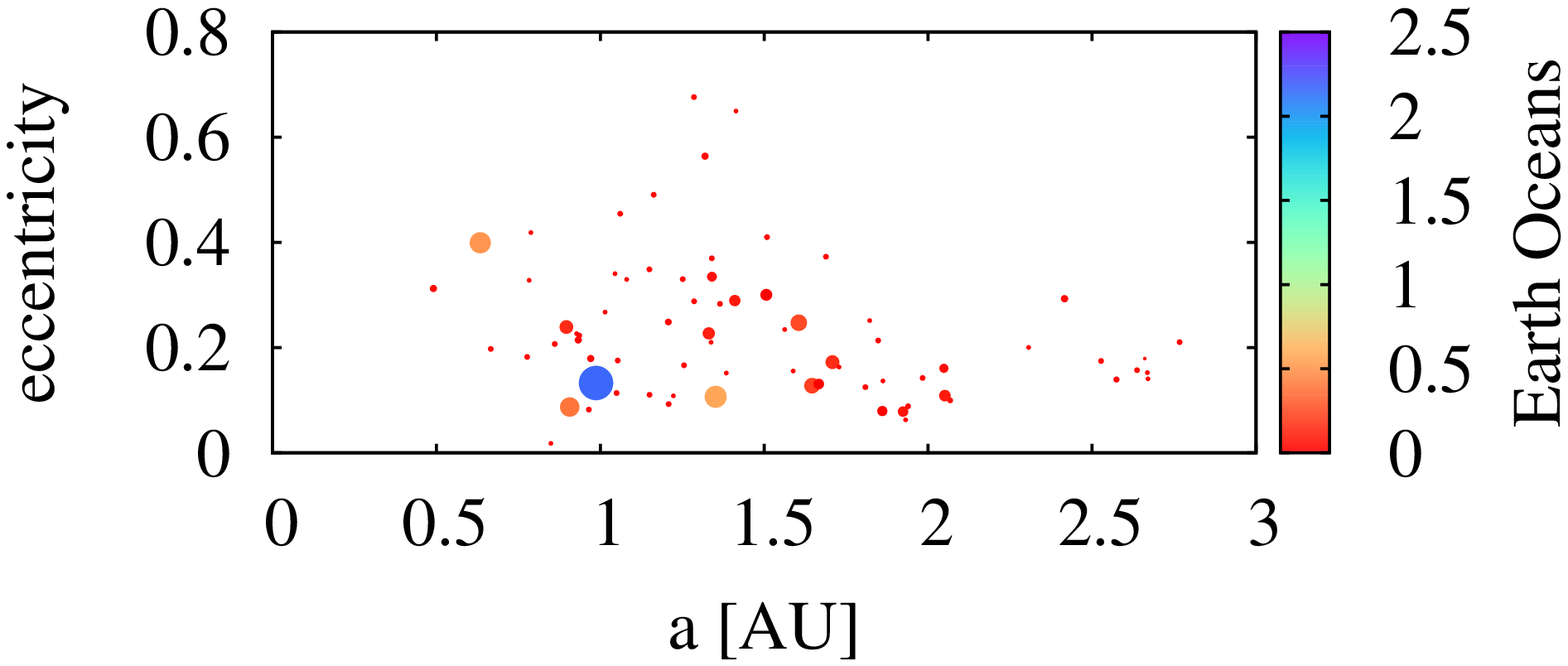}}\qquad
          {\includegraphics[clip,trim=0 2mm 0 7mm,width=0.45\linewidth]{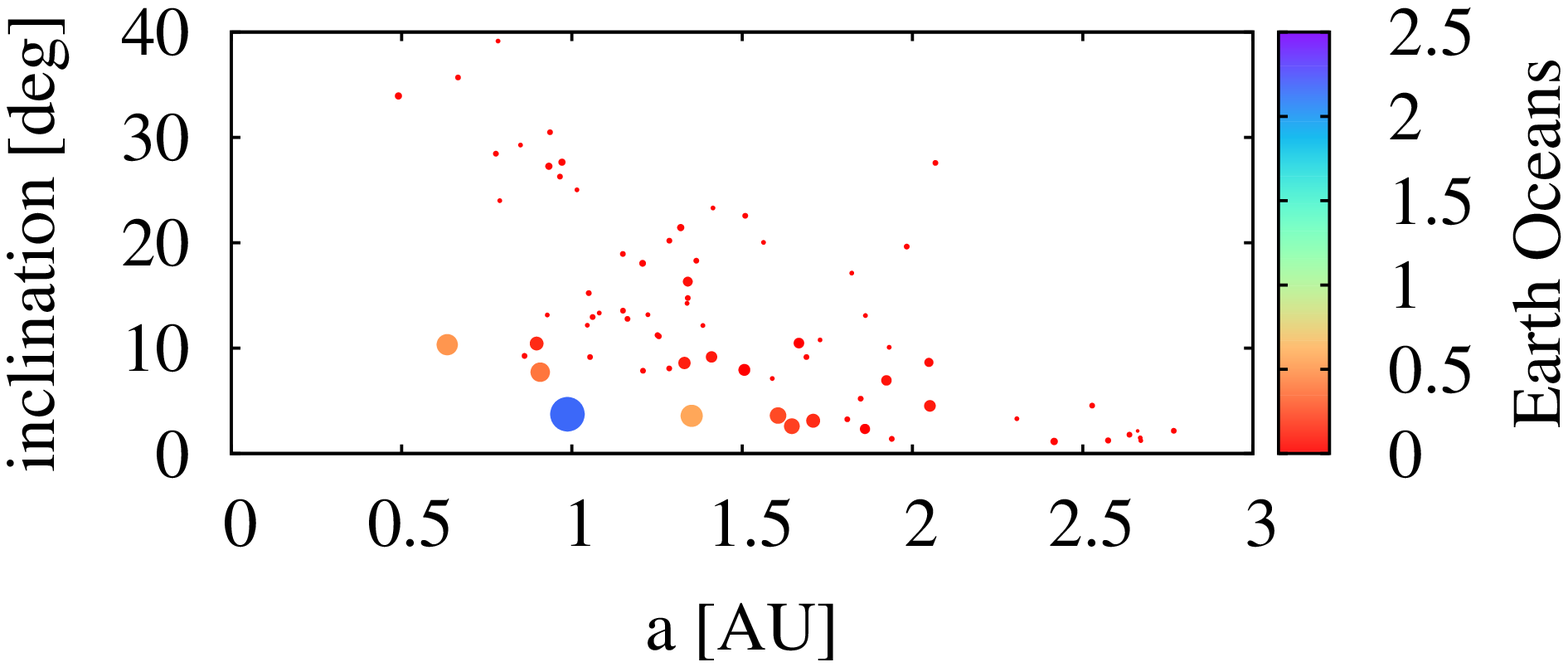}}
        \end{center}
\caption{Eccentricities and inclinations for a typical scenario. The water content is color-coded, %Different colors indicate the water contents; 
circle sizes denote the bodies' masses after 5\,Myrs. The largest bodies are of $\sim 30$ Lunar masses.}
\label{fig:planets}
\end{figure}

\begin{figure}
\centering
\setlength{\unitlength}{\textwidth}
        \begin{picture}(0.8, 0.24)
\put(0.0,0.12){(a)}
\put(0.05,0){\includegraphics[height=0.24\textwidth]{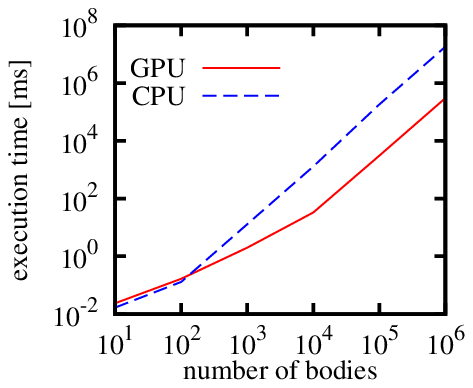}}
\put(0.45,0.12){(b)}
\put(0.5,0.02){\includegraphics[angle=90,height=0.20\textwidth,width=0.2\textwidth]{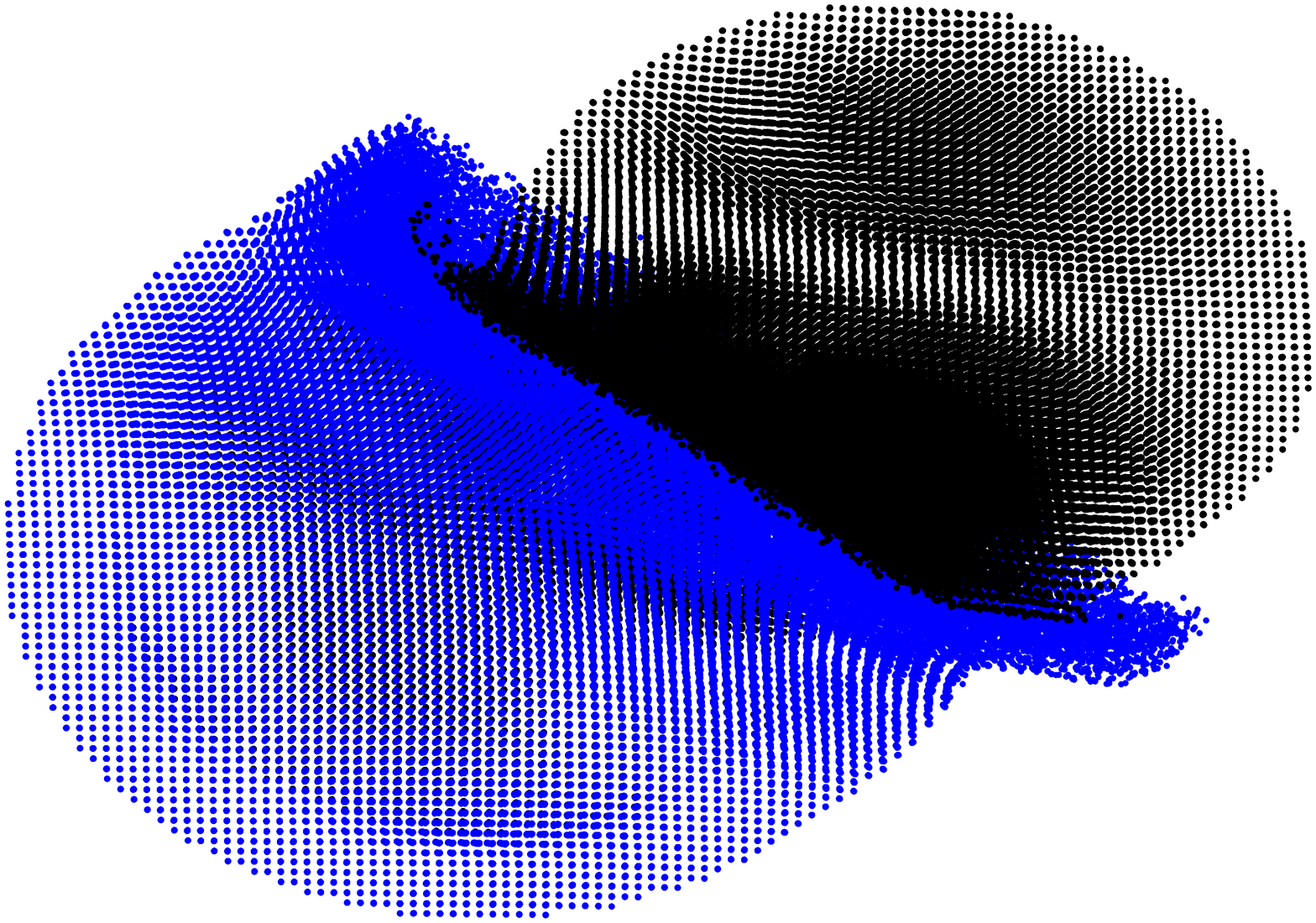}}
        \end{picture}
\caption{a) The GPU-over-CPU computational gain first grows with the number of bodies involved and settles at a factor of approximately 100. b) Low-resolution SPH simulation of two Ceres-mass bodies colliding at 1\,km/s and a $30^\circ\/$ impact angle (50k SPH particles). Blue dots: 30\,wt-\% water ice mantle on the target, black dots: solid basalt projectile.}
\label{fig:gpu}
\end{figure}

Figure~\ref{fig:planets} shows a typical result out of 30 different runs after an integration time of 5\,Myrs using the Lie integration method with adaptive step size \citep{eggdvo10}. We are able to include thousands of gravitating bodies instead of hundreds by using our new massively parallelized GPU code \citep[a descendant of the one described in][]{sul13} which is up to two orders of magnitude faster than our CPU n-body integrator (cf.\ Fig.~\ref{fig:gpu}a). This large number of bodies will take care of {dynamical friction} with a precision not achieved up to now \citep[e.g., ][]{obrmor06}.

\section{Conclusions}

While there exist several results from different groups with different underlying assumptions \citep[e.g., ][]{izihag14,han09}, we believe that {independent computations} %(with others than the usual codes)
including phenomena like dynamical friction and realistic collisional water transfer are desirable for answering the key questions of terrestrial planet formation.

\begin{acknowledgements}
We thank for support from the FWF Austrian Science Fund project S~11603-N16.
%This publication was supported by FWF. 
%In part the calculations for this work were performed on the hpc-bw-cluster---we gratefully thank the bwGRiD project\footnote{bwGRiD (http://www.bw-grid.de), member of the German D-Grid initiative, funded by the Ministry for Education and Research (Bundesministerium fuer Bildung und Forschung) and the Ministry for Science, Research and Arts Baden-Wuerttemberg (Ministerium fuer Wissenschaft, Forschung und Kunst Baden-Wuerttemberg).} for the computational resources.
\end{acknowledgements}

\bibliographystyle{aa} % style aa.bst
\bibliography{references} % references.bib

\end{document}